# PHYSICS

# Interface inter-atomic electron-transition induced photon emission in contact-electrification

Ding Li[1,2,3†], Cheng Xu[1,4†], Yanjun Liao[1,2], Wenzhe Cai[1,4], Yongqiao Zhu[1,4], Zhong Lin Wang[1,2,3,5,6]*

Contact electrification (CE) (or triboelectrification) is the phenomena where charges are produced through physical contact between two materials. Here we report the atomic featured photon emission spectra during CE between two solid materials. Photon emission provides the evidence that electron transfer takes place at the interface from an atom in one material to another atom in the other material during CE. This process is the contact electrification induced interface photon emission spectroscopy (CEIIPES). It naturally paves a way to a spectroscopy corresponding to the CE at an interface, which might impact our understanding of the interaction between solids, liquids, and gases. The physics presented here could be expanded to Auger electron excitation, x-ray emission, and electron emission in CE for general cases, which remain to be explored. This could lead to a general field that may be termed as contact electrification induced interface spectroscopy (CEIIS).

## INTRODUCTION

Contact electrification (CE), which is the scientific term for the well-known triboelectrification, means that the charges are produced because of physical contact. CE is a universal phenomenon in both our daily life and nature world (*1–6*). When we walk along a road, there are CEs between shoes and the ground. When we wash our cars, there are CEs between the car shell and water. When the clouds move in air, there are CEs between clouds and air. When the earth shakes, there are CEs between the interfaces of rocks. Although CE was recorded first as early as 2600 years ago during ancient Greek civilization, the mechanism behind CE is still on debate regarding to if CE is due to electron transfer (*7*, *8*), ion transfer (*9*, *10*), or even materials species transfer (*11*, *12*). The research on the mechanism of CE recently enormously evolved with modern technologies (*13*, *14*). With some unexplainable even contradictory observations (*15–19*), it seems that no one mechanism can explain such a complex phenomenon (*20*). Is there really no way that we could access the fundamental or dominant mechanism to generalize the CE concept for different types of materials?

In 2012, Wang and colleagues (*21*) invented triboelectric nanogenerator (TENG) and shed light on mechanical energy convertor into electric power with applications (*22*) as micro/nanopower sources, self-power sensing, blue energy, and high-voltage power sources. Moreover, it serves as a unique probe to explore the basic mechanism behind CE into great detail by a direct measurement of the surface charge density. In 2018, Wang and colleagues (*23*) explored the temperature-dependent real-time charge transfer in CE by TENG and pointed out that electron transfer is the dominant process for CE between metal and ceramic. Then, they further revealed that the dominant deterring factor of CE at high temperatures is the electron thermionic emission (*24*) and developed electron cloud potential well model for understanding CE (*25*, *26*). The electron transfer mechanism behind CE was further confirmed at nanoscale using Kelvin probe force microscopy that it happens at repulsive region when two atoms are close to each other (*27*). They also conducted detailed analyses from several aspects, such as photon excitation effect (*28*), atmosphere effect on the surface states of dielectrics (*29*), surface functional groups on CE at liquid-solid interface (*30*), tribovoltaic effect at the interface of semiconductors (*31*), and quantifying electron transfer in liquid-solid interface with a two-step model for electric double layer (*32*). On the basis of above experimental results as supported by quantum mechanical calculations (*33*, *34*), Wang and Wang (*35*) propose that an interatomic interaction model for general CE cases. Wang predicted that in the process of electron transfer, as energy is dissipated, there must be characteristic photons emission from atomic outer shell related to CE, which could give birth to a new optical spectroscopy for studying electronic transitions at interfaces. Unfortunately, this may be a forgotten field in the long history of CE studies because of the complication from air discharge and the weak photon signals.

Here, we observed atomic featured photon emission spectra during CE at a solid-solid interface by contacting fluorinated ethylene propylene (FEP) with acrylic or FEP with quartz. The physical processes of typical photon lines, such as H atom lines (486 and 656 nm) and O atom lines (715, 799, and 844 nm) from quartz interface and F atom lines (782 and 760 nm) from FEP interface, were identified. They are associated with electron transitions through energy resonance transfer between two atoms from different materials when they are so close that they are in the repulsive force region. Unlike triboluminescence (*36*, *37*) reported previously with association to air discharge or the breakdown of chemical/ionic bonds under stress or fraction, the characteristic photon emission induced by CE carries abundant information about the energy structure at the interfaces. Three possible physical processes are suggested for understanding the photon emission arising from the electron charge transferred in CE: (i) transition to a lower energy level in one atom by emitting a photon, (ii) transition to the excited state of another atom through energy resonance transfer, followed by transiting to a lower energy level of the atom, and (iii) direct transition to a lower energy level in another atom, followed by transiting to an even lower energy level by a photon.

[1]CAS Center for Excellence in Nanoscience, Beijing Key Laboratory of Micro-nano Energy and Sensor, Beijing Institute of Nanoenergy and Nanosystems, Chinese Academy of Sciences, Beijing 101400, China. [2]Center on Nanoenergy Research, School of Physical Science and Technology, Guangxi University, Nanning 530004, China. [3]School of Nanoscience and Technology, University of Chinese Academy of Sciences, Beijing 100049, China. [4]School of Materials Science and Engineering, China University of Mining and Technology, Xuzhou 221116, China. [5]CUSTech Institute, Wenzhou, Zhejiang 325024, China. [6]School of Materials Science and Engineering, Georgia Institute of Technology, Atlanta, GA 30332-0245, USA.
*Corresponding author. Email: zhong.wang@mse.gatech.edu
†These authors contributed equally to this work.

This is the CE-induced interface photon emission spectroscopy (CEIIPES) for studying electronic transitions at solid-solid interfaces.

## RESULTS
### Working principle of CE setup
The experimental setup is illustrated in Fig. 1. The core parts were composed of a hollow cylinder sandwiched between a metal cover and a metal base. Inside this cylinder, there were four metal fans driven by a motor. Materials for CE could be either attached to metal fans or to the cylinder. When the fans were rotated, CE occurred at the interface. To minimize air discharge effects, we placed the core parts in a vacuum chamber (31 cm in diameter and 50 cm in height) with a vacuum pump to achieve an atmosphere with controlled pressure. In addition, the pressure was measured by a pressure meter and controlled by the differential flow of inlet and outlet of the vacuum chamber through glass tube float flow meters. There was an observatory opening on this chamber with a quartz window, which is transparent from deep ultraviolet to infrared. If there were photon signals coming from the core parts, then they would be recorded by the spectrometer with a sensitive charge-coupled device (CCD) detector, as in Fig. 1A. There were two measurement modes in our experiments. Mode A, contact-sliding mode, was a material attached to metal fans and directly contacted with the inner side wall of the cylinder. For example, in Fig. 1B, the FEP was attached to metal fans and directly contacted with cylinder made of acrylic. When metal fans were rotated, CE occurred between the interface of FEP and acrylic in sliding mode. Mode B, contact-separation mode, was one material attached to metal fans and another material attached to the inside wall of cylinder with tiny gap (<1 mm). The vertical-view image of this mode is in Fig. 1C, where copper foil was attached to the metal fans, while FEP was attached to the inside wall of quartz cylinder as an example. The schematic figure of this mode was in Fig. 1E, where nylon was attached to the metal fans, while FEP was attached to the inside wall of the quartz cylinder as an example. In addition, the working principle of Mode B is demonstrated in Fig. 1D. When a metal fan was rotated, it contacted the surface of FEP on and off alternatively. When it was on the surface of FEP, it would compress the FEP to contact with the inside wall of quartz cylinder. Electrons transition at the interface due to CE occurred simultaneously. When it was off the surface of FEP, it would release the FEP and separate the FEP from quartz cylinder automatically, leaving negative charges on FEP and positive charges on quartz.

### Photon emission during CE
We have carefully excluded the features contributed by air discharge in the measurements by comparing the spectra acquired at different pressures (more detail in fig. S1), so that our data are focused on the photon emission from the interface during CE. For CE of the FEP-acrylic group at pressure around 24 Pa, FEP on metal fans with acrylic cylinder in mode A driven by a motor, several sharp lines were observed with atomic spectra feature such as discrete distribution in the spectrum and very narrow (<1 nm) full width at half maximum (FWHM) as in Fig. 2A. According to their peak positions, we identified some main peaks. We attributed the peak positions at 434.0, 486.0 (inset of Fig. 2A), and 656.2 nm to the electron transitions in hydrogen (H) atom (red) and those at 777.5 and 844.7 nm to the electron transitions in oxygen (O) atom (magenta). It was quite out of our expectations that we could observe the atomic spectra of H atom. Therefore, we further used the high-resolution grating of 1800 lines/mm to recorded 486.0- and 656.2-nm lines with better accuracy to confirm. In Fig. 2B, the peak position was 486.17 nm in experiments with FWHM of 0.192 nm. In addition, the corresponding standard H line is 486.135 nm (*38*). Then, the difference between them is 0.034 nm, which is in the range of uncertainty for our spectrometer. It corresponds to the photon emission due to the electron transition from the excited state $n = 4$ to the excited state $n = 2$ in H atom. In Fig. 2C, the peak position was 656.34 nm in experiments with FWHM of 0.209 nm, and the corresponding standard H line is 656.285 nm (*38*). Then, the difference between them is 0.058 nm, which is also in the range of uncertainty for our spectrometer. It corresponds to the photon emission due to the electron transition from the excited state $n = 3$ to the excited state $n = 2$ in H atom. To our surprise, we even observed weak signals of photo emission corresponding to the electron transition from the excited state $n = 5$ to the excited state $n = 2$ in H atom (434.0 nm).

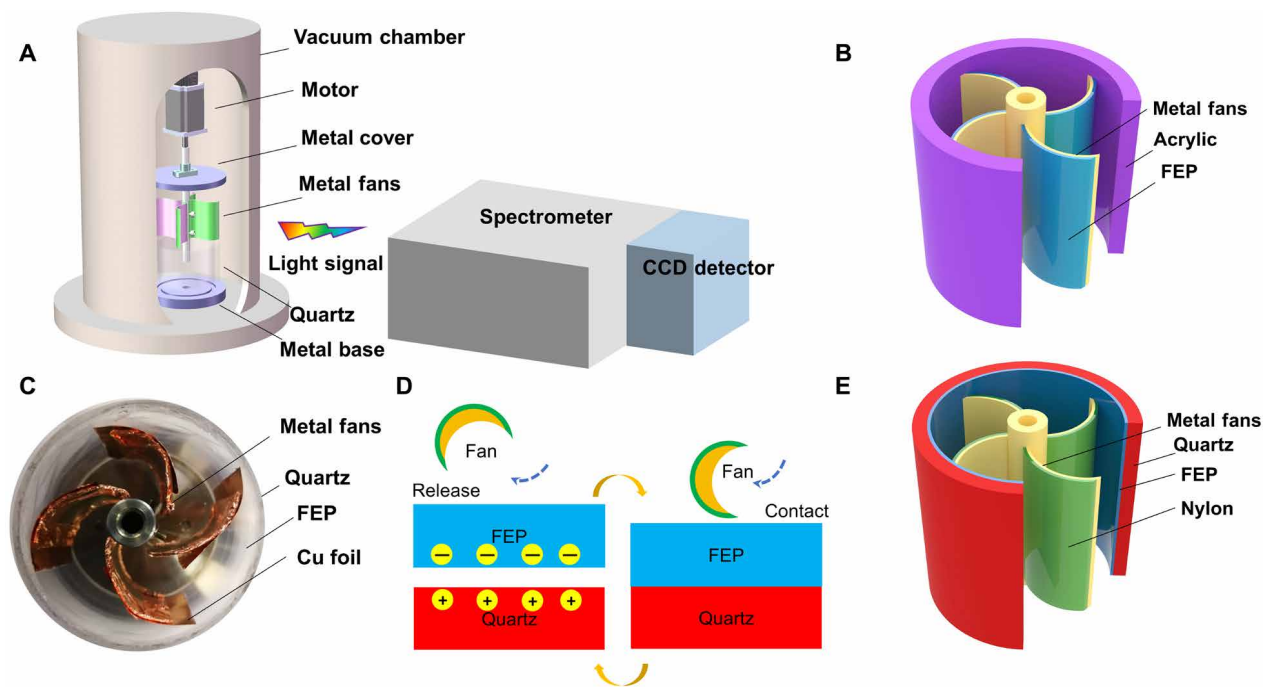

**Fig. 1. The schematic diagram of experiments.** (**A**) CE parts in vacuum chamber and spectrometer. (**B**) Measurement mode A: FEP on fans and directly contact with quartz or acrylic. (**C**) Optical photograph of the CE parts. (**D**) The working principle of the measurement mode B. (**E**) Measurement mode B: FEP attached to quartz or acrylic with nylon, etc. on fans. Photo credit: Ding Li, Beijing Institute of Nanoenergy and Nanosystems, Chinese Academy of Sciences.

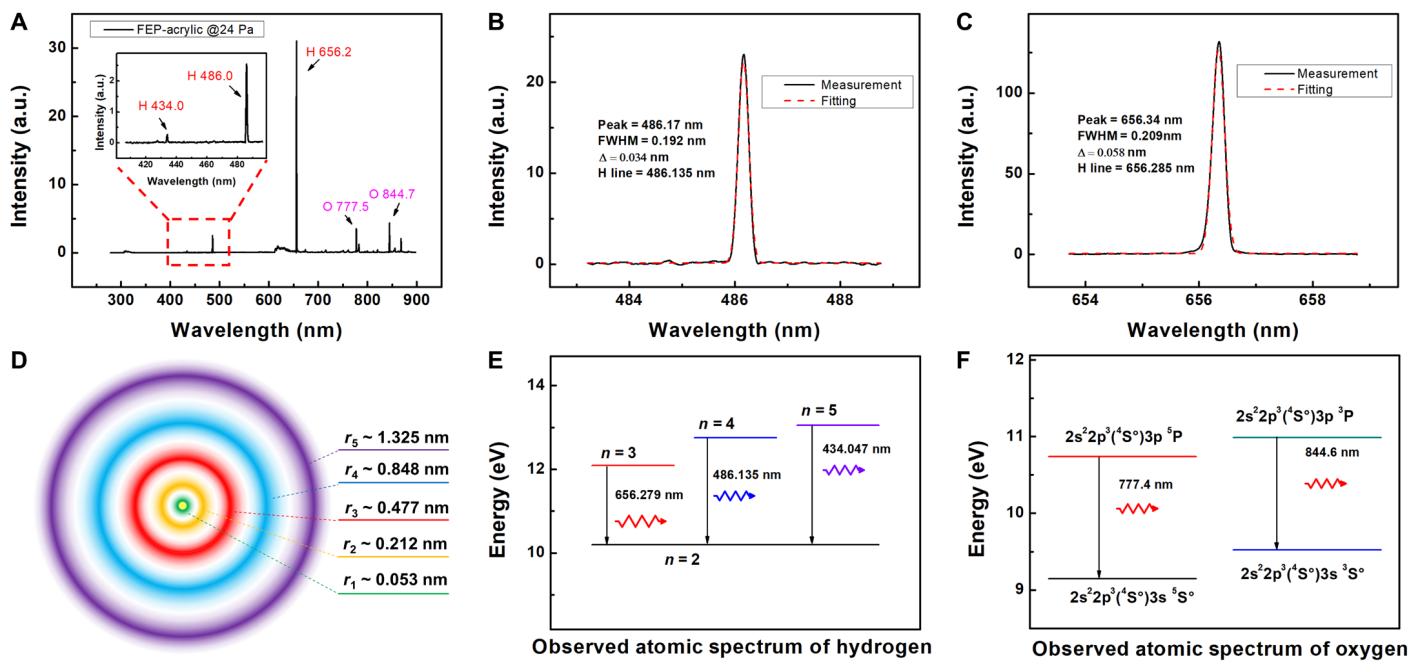

**Fig. 2. Interface electron transition induced photo emission spectra and related energy levels in CE at low pressure for the FEP-acrylic group.** (**A**) The spectra recorded at 24 Pa with identified hydrogen and oxygen atomic spectra. a.u., arbitrary units. (**B** and **C**) For hydrogen spectra, higher-resolution grating was used for further confirmation. (**D**) Electron energy radius on Bohr model of hydrogen atom. (**E** and **F**) Energy levels for identified atomic lines in (A).

The significant contribution of H to the CEIIPES could be understood from the consideration of atom size during CE. For simple estimations, we use the Bohr model of hydrogen atom on the electron energy radius (*39*)

$$r_n = \frac{4\pi\varepsilon_0 \hbar^2}{m_e e^2} \cdot n^2 \quad (1)$$

where $e$ and $m_e$ are constancies, representing electronic charge and mass of electron. $\hbar = h/2\pi$ where $h$ is Planck's constant. $r_n$ and $n$ represents Bohr radius of hydrogen atom for $n$ energy level and the number $n$ energy level, respectively. The radii for the states of $n = 5$, 4, 3, 2, 1 are 1.325, 0.848, 0.477, 0.212, and 0.053 nm, respectively (Fig. 2D). These large atomic radii made electron wave function of H atom at excited states easy for overlapping with those of other atoms during CE, which is a physical process by compressing one material against the other, so that the chance for electrons transferring from one atom to H atom or from H atom to another would increase. The energy levels for electron transitions we observed of H atoms are summarized in Fig. 2E. In addition, the energy levels for electron transitions we observed of O atoms at this pressure are summarized in Fig. 2F. The photon emission around 777 nm corresponds to the electron transition from $2s^22p^3(^4S°)3p$ $^5P$ to $2s^22p^3(^4S°)3s$ $^5S°$, and the one around 844 nm corresponds to the electron transition from $2s^22p^3(^4S°)3p$ $^3P$ to $2s^22p^3(^4S°)3s$ $^3S°$ (*38*, *40*, *41*).

As predicted by Wang (*42*), there would be photon emission associated with the physical processes of CE. In addition, these photon emissions with atomic spectra features observed could associate with electron transitions during CE, which is thus defined as the contact electrification induced interface photon emission spectroscopy CEIIPES.

Figure 3 (A and D) depicts the CEIIPES from CE of the FEP-acrylic group at different atmosphere pressures, from the deep ultraviolet to near infrared (from 277 to 897 nm). At pressure below 92 Pa, only several dominant lines are observed, and 656.2-nm line is corresponding to the transition from $n = 3$ to $n = 2$ of the H atom. When the pressure increases, more atomic spectra featured lines appeared, and the intensity of 656.2-nm line decreases. In addition, the strong lines in these spectra are mostly distributed in the range between 600 and 900 nm. For more detailed analysis, we take the one at 200 Pa for example and enlarge the spectrum range where strong lines distributed (Fig. 3B). According to the Atomic Spectra Database provided by National Institute of Standards and Technology (*38*), we identify these strong lines to different elements in our experiments (more detail information in note S3). We denote lines with wavelength we observed and colors for hydrogen atom with red, carbon atom with blue, oxygen atom with magenta, fluorine atom with black, and unidentified with green. In addition, the peak intensities of typical H atom lines (486.1 and 656.3 nm) and O atom lines (777.5 and 844.7 nm) changing with pressure are depicted in Fig. 3E. For H atom lines, the intensities are the strongest at 24 Pa and

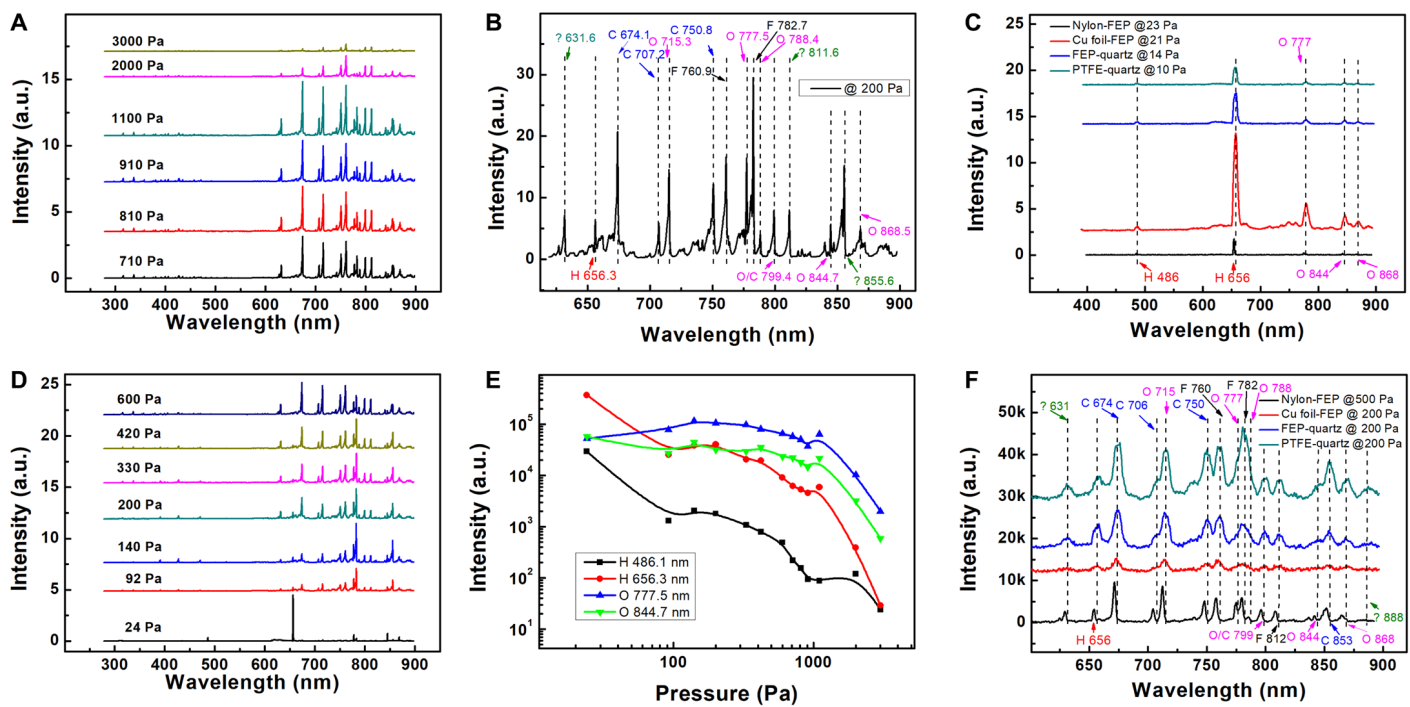

**Fig. 3. Interface electron transition induced photo emission spectra and related energy level in CE at different pressures for different contact materials groups.** (**A** and **D**) CEIIPES of the FEP-acrylic group at different atmosphere pressures. (**B**) Enlarge and identifications of atomic lines in CEIIPES of the FEP-acrylic group at 200 Pa. (**C** and **F**) CEIIPES of different groups at different atmosphere pressures with identifications of atomic lines. (**E**) The peak intensity of selected atomic lines changes with atmosphere pressure.

then decrease with pressure with slightly increasing around 200 Pa and markedly decreasing beyond 1000 Pa. For O atom lines, the intensities initially increase from 24 to 200 Pa at the maximum and then gradually decrease with further markedly decreasing beyond 1000 Pa. If the surface adsorbates really play a key role here, the emission intensity would increase as a function of ambient pressure. In practice, the intensity of CEIIPES decreases with pressure. Although, we could not explain well about the reasons for the intensity changes with pressure now, we will make quantitative studies in the near future. Our focus here is about the physics process regarding to charge transfer between the two materials.

These atomic spectrum featured photon emission lines induced by CE are not only observed in FEP-acrylic group, FEP-quartz group, and polytetrafluoroethylene-quartz (PTFE-quartz) group at mode A but also exist in other groups at mode B, such as nylon-FEP-quartz group and Cu foil-FEP-quartz group, as shown in Fig. 3C (pressure below 25 Pa) and Fig. 3F (pressure above 200 Pa). In addition, the signal intensities are different for different groups. We tried to get signals as clear as possible by changing slits and gratings of spectrometer. Despite the different measurement parameters, H atom lines (486 and 656 nm) and O atom lines (777 and 844 nm) are still visible for pressure below 25 Pa with the dominant line of 656 nm. For pressure above 200 Pa, the peak distributions are similar as the ones in Fig. 3B. The signals we observed are reliable considering different measurement parameters and CE groups. It could be possible that CEIIPES is a common effect in CE.

## Physical processes of electron transfer
The physical processes behind these atomic spectrum featured photon emission lines are illustrated in Fig. 4 with respect to the energy levels and electron transitions. Figure 4A depicts schematically the interface of FEP and quartz at atomic level using VESTA (*43*). The H, O, Si, F, and C atoms are represented by yellow, red, gray, blue, and brown balls, respectively. In addition, the quartz is in projection along a-axle. When FEP contacts with quartz, there would be electron transitions for these materials. For instance, electrons transfer could occur between F and O atoms, between F and H atoms, and even between H and O atoms. In addition, photon emissions are accompanied by these transitions. When F and H atoms are close to each other at the repulsive force region, which means that the two have a strong electron cloud overlap, electrons might transfer between these atoms through energy resonance transfer process. For example, the 656.279-nm line represents transition from $n = 3$ level to $n = 2$ level in H atom, and 782.3-nm line represents transition from $2s^22p^4(^3P)3d\ ^2F$ to $2s^22p^4(^3P)3p\ ^4P°$ in F atom. When we consider the vacuum level as 0 eV and the first ionization energy as the energy difference between the vacuum level and the ground state, we could draw them together in energy diagram as in Fig. 4B. It could be immediately found that $n = 3$ energy level of H atom is so close to the $2s^22p^4(^3P)3d\ ^2F$ of F atom that electrons could easily transfer between these levels by the energy resonance transfer process. Since we observe 656-nm line of H atom during CE, it is the evidence that there are electrons occupied the $n = 3$ energy level. Considering that FEP is negatively charged after it contacts with quartz, there should be electron transfer from quartz to FEP. In addition, one possible route for this physical process is through energy resonance transfer from $n = 3$ energy level in H atom to $2s^22p^4(^3P)3d\ ^2F$ of F atom. When it is on $2s^22p^4(^3P)3d\ ^2F$ of F atom, it further transits to the lower energy level $2s^22p^4(^3P)3p\ ^4P°$ of F atom with a photon emission. In addition, the 782-nm lines we observed are the perfect evidence of these transitions.

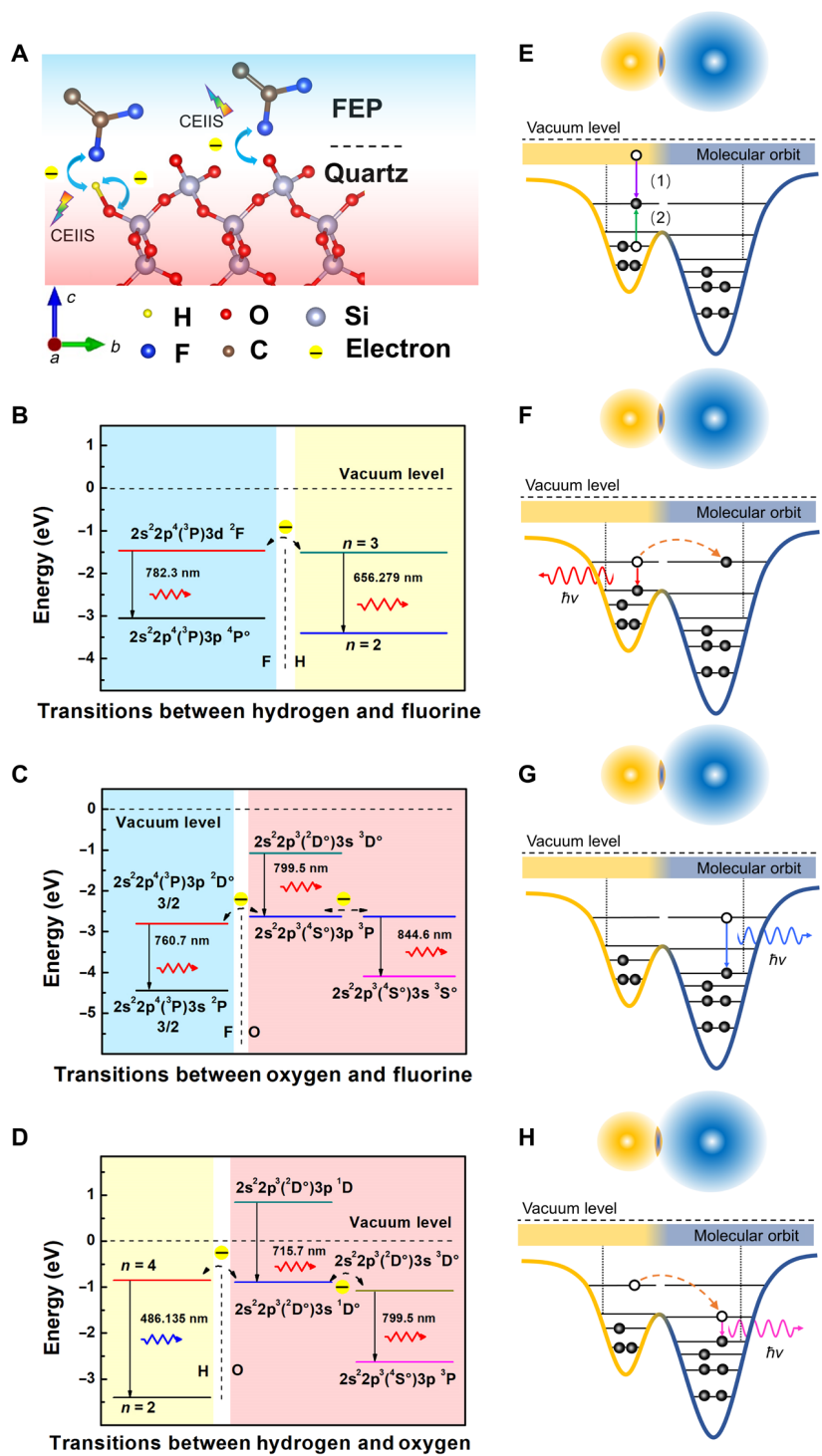

**Fig. 4. Energy diagram for interface electron transition induced photo emission.** (**A**) The schematic diagram of FEP and quartz interface at atomic level. (**B**) Energy diagram of electron transition between hydrogen and fluorine. (**C**) Energy diagram of electron transition between oxygen and fluorine. (**D**) Energy diagram of electron transition between hydrogen and oxygen. In addition, the schematic diagram of possible physical processes of electrons transitions and the associated photon emission, also known as Wang transition (*42*), when two atoms are close to each other (**E** to **H**) (see text for details).

Similar behavior could also be observed when a F atom contacts with an O atom, by aligning their energy levels with vacuum level set as 0 eV (Fig. 4C). High-energy electrons in $2s^22p^3(^2D°)3s\ ^3D°$ could transit to a lower level of $2s^22p^3(^4S°)3p\ ^3P$ in O atom with photon emissions (799-nm line). When the electrons are at level $2s^22p^3(^4S°)3p\ ^3P$ of O atom, they could have two possible choices. One is to further transit to a lower energy $2s^22p^3(^4S°)3s\ ^3S°$ in O atom with photon emissions (844-nm line). It is also the evidence that there are electrons at level $2s^22p^3(^4S^o)3p\ ^3P$ of O atom. Another is to transit to the $2s^22p^4(^3P)3p\ ^2D^o$ 3/2 of F atom by energy resonance transfer during CE. After that, it further transits to the lower energy level $2s^22p^4(^3P)3s\ ^2P$ 3/2 of F atom with a photon emission (760-nm line). Considering that FEP is negatively charged after it experiencing CE with quartz, there should be electron transition from O atom to F atom. In addition, the 760-nm lines observed are the evidence of this physical process.

During CE between FEP and quartz, there are also electron transitions between H atom and O atom at the surface of quartz as in shown in Fig. 4D. Electrons at a higher energy level [$2s^22p^3(^2D^o)3p\ ^1D$ of O atom] transit to a lower level [$2s^22p^3(^2D^o)3s\ ^1D^o$ in O atom] with photon emissions (715-nm line). When the electrons are at level $2s^22p^3(^2D^o)3s\ ^1D^o$ of O atom, they could transfer to another energy level through energy resonance transfer with two possible choices. One is to $2s^22p^3(^2D^o)3s\ ^3D^o$ energy level, and it could further transit to a lower energy $2s^22p^3(^4S^o)3p\ ^3P$ in O atom with photon emissions (799-nm line). Another one is to transit to $n = 4$ energy level of H atom by energy resonance transfer during. After that, it further transits to the lower energy level $n = 2$ of H atom with a photon emission (486-nm line). In addition, the lines we observed are evidence of these physical processes.

As discussed above, the possible physical process routes for electron transitions among different atoms during CE are schematically summarized in Fig. 4 (E to H), taken atom A (yellow) and atom B (blue) for example. When atom A and atom B are pressed close to each other in the repulsive force region, the energy potential barrier between the two is lowered owing to strongly electron wave functions overlap. Inside the energy potential wells of each atom, there are energy levels possible for electrons to occupy distributing from the ground state to vacuum level. The closer to the vacuum level, the more energy levels are concentrated. We use dotted lines to represent this highly concentrated energy level close to the vacuum level and denote some of them in our schematic figures. During the CE, some of electrons are temporarily transmitted to excited states. In addition, there are two possible ways for electrons transit to excited states (Fig. 4E): (i) The electron transit from molecular orbit to the excited state of an atom, during which it might experience nonradiative decay or radiative decay; (ii) the electron is excited from lower energy level to higher energy level inside an atom. When the electron is at excited state, it could transit to a lower energy level by emitting a photon. It could also transit to the excited state of another atom (from atom A to atom B) through energy resonance transfer if the energy level of excited state in atom A is close to the one of another excited state in atom B (Fig. 4F). Then, the electron, transferring from atom A, could transit to a lower energy level in atom B by emitting a photon (Fig. 4G). It is also possible, as revealed by previous theoretic analysis, that electron in atom A with a higher energy level could transit to a lower energy level in atom B. It is followed by a transition to an even lower energy level accompanied by a photon emission (Fig. 4H).

CEIIPES is different from the fluorescence spectra for molecules. First, the origins of photon emission are different. CEIIPES is a photon emission associated with electron transfer between two atoms, while fluorescence spectra are associated with electron transition between molecular levels with many vibrational sublevels. Second, the spectra features are different. CEIIPES is discrete atomic featured sharp lines with FWHM less than 1 nm, while fluorescence spectra are normally continuous board bands with FWHM up to tens of nanometers (fig. S2).

### The role of H atom in CE

From our experiments, we could speculate that H atom may have unique roles in CE. Comparing the intensities of 656-nm line of FEP-quartz group with FEP-acrylic group in mode A, the intensities were proportional to the H atom density at the interface (Fig. 5, A and C). For the FEP-acrylic group, the intensity of 656-nm line was 451,355 counts with a slit width of 100 μm, while the one in the FEP-quartz group was 19,374 counts with a slit width of 350 μm. For the sample line, the intensity with slit width of 100 μm is half of the one with a slit width of 350 μm (fig. S1F). Hence, the ratio of the intensity of 656-nm line for FEP-quartz group and FEP-acrylic group was about $2 \times 10^{-2}$. In quartz, the $OH^-$ content was $148.2 \times 10^{-6}$ g in 1 g of quartz according to the results from National Safety Glass and Quartz Glass Quality Supervision and Inspection Center of China, which corresponded to around $5 \times 10^{19}$ H atom in 1 mol quartz (Fig. 5B). In addition, it is $8 \times 10^{21}$ H atom in 1 mol acrylic (Fig. 5D) according to its molecular formula. At the interface, the ratio of H atom for quartz and acrylic was also about $2 \times 10^{-2}$. It is the evidence that 656-nm line is proportional to the density of H atoms at the interface. The stronger the intensity is, the more electrons transfer would be.

According to Eq. 1, H atom holds the largest Bohr radius for excited states of electron among all the basic chemistry elements. Since H atom has the largest Bohr radius for excited states, it would be easier to overlap electron wave functions with other atoms in space. Take the interface of FEP and quartz for example and compare F-O interface and F-H-O interface. If electrons transfer between F atom and O atom with H atom as media in between, then it would consume less energy than the one at the F-O interface. Since H atom has the largest Bohr radius for excited states, the distant need for F-H-O in repulsive region is also longer than the one for F-O interface. Therefore, less energy is needed for bring them in repulsive region. Although H atom has the largest Bohr radius for excited states, which is easier for approaching other atom in repulsive region, it has only several energy levels, and the chances for energy resonance transfer are relatively low. However, O atom has abundant energy levels, and the chances for energy resonance transfer are relatively high, as in Fig. 5C. It might help us to understand that O is usually very active in chemical reactions.

Nevertheless, we only discussed a few strong lines that can be identified from CEIIPES, and there might be other important signals, which might be too weak to distinguish from noise. Because of the limitations our spectrometer, we could not further resolve the CEIIPES with higher resolution and more efficient spectrometer. Furthermore, the other possible physical processes, such as vibrational, rotational, and even spin related phenomenon, are not discussed. As the signals of CEIIPES are really very weak in our experiments, we have to prolong the detection time to get relatively reliable spectra. At present, we report photon emission related to CEIIPES only at solid-solid interfaces. We speculate that there would be more interesting phenomena revealed by CEIIPES at solid-liquid, solid-gas, gas-gas, gas-liquid, and liquid-liquid interfaces.

### DISCUSSION

Here, we observed atomic featured photon emission spectra during CE between two solids. The photon emission is the evidence that electrons transfer takes place from one atom in one material to another atom in another material at the interface during CE. This process is the CEIIPES. Three possible physical processes are suggested for understanding the photon emission arising from the electron charge transferred in CE: CE induced electron to transit (i) to a lower energy level in one atom by emitting a photon; (ii) to the excited state of another atom through energy resonance transfer, followed by transiting to a lower energy level of the atom; and (iii) to a lower energy level in another atom, followed by transiting to an even lower energy level by a photon. CEIIPES occurs through energy resonance transfer when atoms from different materials are brought close with each other. Clarifying physical processes behind CE, it helps us better understand how two materials are charged after CE. It naturally paves a way to a spectroscopy corresponding to the CE at an interface, which might have fundamental impacts to understand the interaction among solids, liquids, and gases. Although we only focused on photon emission in CE for solid-solid case in this study, it could be expanded to Auger electron excitation, x-ray emission and electron emission in CE for general cases, which remain to be explored. This could be a general field that may be termed as contact electrification induced interface spectroscopy (CEIIS).

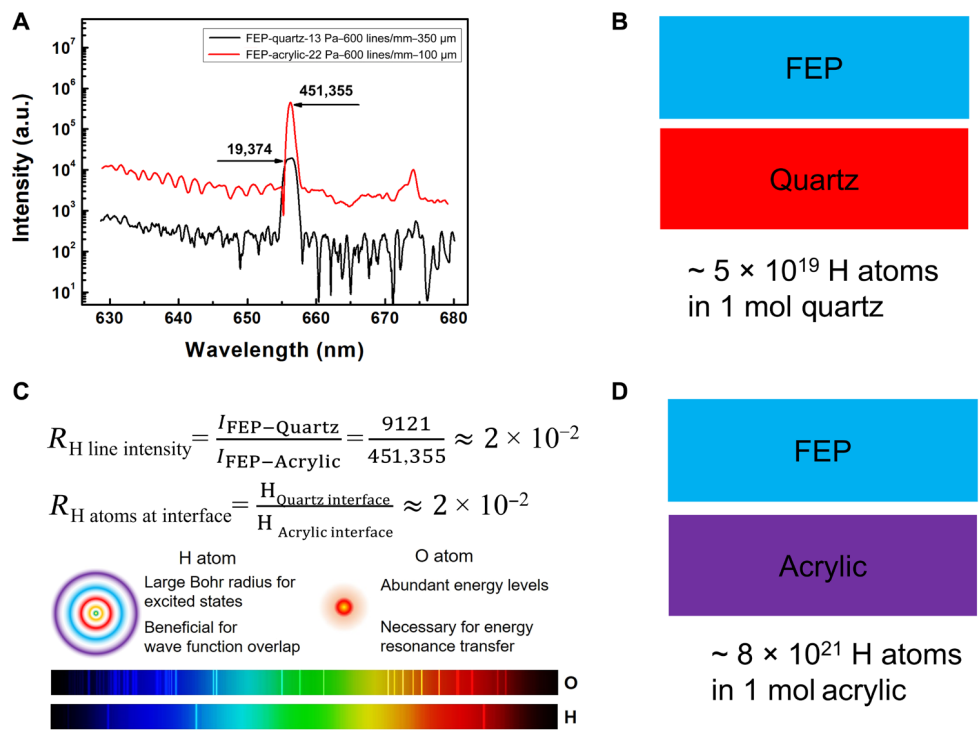

**Fig. 5. Interface electron transition induced photo emission intensity is comparable to the H atoms at the interfaces for FEP-acrylic group and FEP-quartz group.** (**A**) Take the H 656.2-nm line for example and the corresponding illustrations in (**B**) and (**D**). (**C**) Color spectra of elements H and O in the range of 400 to 700 nm, showing different functions of them for electron transfer at CE. The ratio of intensity is comparable to the ratio of H atoms at the interfaces.

## MATERIALS AND METHODS
### The setup for CEIIPES observation
The speed of metal fans was control by the motor at 200 rpm, and the diameter of the quartz or acrylic was 10 cm with a height of 8 cm. The FEP, quartz (transparent in deep ultraviolet region), acrylic, and Cu foil were all commercially available. The spectrometer was Andor Kymera 328i equipped with EMCCD Newton DU971P-UVB (enhance mode), the grating of 600 lines/mm blazing at 500 nm and holographic grating of 1800 lines/mm. The slit was usually 100 or 350 μm. For slit of 100 μm, the resolution of our spectrometer was 0.21 nm for grating of 600 lines/mm and 0.06 nm for grating of 1800 lines/mm. Since the photon signals were really very weak, the spectra ranging from 277 to 897 nm were measured using grating of 600 lines/mm, which were stitched together by six sections (277 to 401 nm, 378 to 501 nm, 479 to 600 nm, 579 to 699 nm, 680 to 798 nm, and 781 to 897 nm). In addition, the exposure time for each section is kept at 10 s with an accumulation of 12 times. For spectra with grating of 1800 lines/mm, they are also recorded at the exposure time of 10 s with an accumulation of 12 times.

### The force measurement
The force gauge we used is M5-10 made by Mark-10 Corporation, USA. The normal force between FEP and the inner side wall of the cylinder could be adjusted by the stiffness and the curvature of metal fans. In our experiments, the normal force is in the range of 0.3 and 6.6 N.

### The pressure measurement of the vacuum system
The pressure meter in our experiments is ZDF-5227AX Composite Vacuum Gauge with ZJ-52T resistance vacuum gauge, both made by Chengdu Reborn Electronics Co. Ltd. with a range of $1 \times 10^5$ to $1 \times 10^{-1}$ Pa. In addition, the diameter of quartz window is 9 cm.

## SUPPLEMENTARY MATERIALS
Supplementary material for this article is available at https://science.org/doi/10.1126/sciadv.abj0349

## REFERENCES AND NOTES

1. W. Jamieson, The electrification of insulating materials. *Nature* **83**, 189 (1910).
2. E. W. B. Gill, Frictional electrification of sand. *Nature* **162**, 568–569 (1948).
3. E. W. B. Gill, G. F. Alfrey, Electrification of liquid drops. *Nature* **164**, 1003 (1949).

SCIENCE ADVANCES | RESEARCH ARTICLE4. J. T. Davies, Measurement of contact potentials at the oil–water interface. *Nature* **167**, 193–194 (1951).
5. E. J. Workman, S. E. Reynolds, Production of electric charges on water drops. *Nature* **169**, 1108–1109 (1952).
6. J. Latham, C. D. Stow, Electrification of snowstorms. *Nature* **202**, 284–285 (1964).
7. C. Liu, A. J. Bard, Electrons on dielectrics and contact electrification. *Chem. Phys. Lett.* **480**, 145–156 (2009).
8. S. Pan, Z. Zhang, Triboelectric effect: A new perspective on electron transfer process. *J. Appl. Phys.* **122**, 144302 (2017).
9. L. S. Mccarty, G. M. Whitesides, Electrostatic charging due to separation of ions at interfaces: Contact electrification of ionic electrets. *Angew. Chem. Int. Ed.* **47**, 2188–2207 (2008).
10. A. F. Diaz, D. Wollmann, D. Dreblow, Contact electrification: Ion transfer to metals and polymers. *Chem. Mater.* **3**, 997–999 (1991).
11. J. Lowell, The role of material transfer in contact electrification. *J. Phys. D Appl. Phys.* **10**, L233–L235 (1977).
12. H. T. Baytekin, B. Baytekin, J. T. Incorvati, B. A. Grzybowski, Material transfer and polarity reversal in contact charging. *Angew. Chem. Int. Ed.* **51**, 4843–4847 (2012).
13. S. Pan, Z. Zhang, Fundamental theories and basic principles of triboelectric effect: A review. *Friction* **7**, 2–17 (2019).
14. Z. Zhang, N. Yin, Z. Wu, S. Pan, D. Wang, Research methods of contact electrification: Theoretical simulation and experiment. *Nano Energy* **79**, 105501 (2021).
15. Y. Fang, L. Chen, Y. Sun, W. P. Yong, S. Soh, Anomalous charging behavior of inorganic materials. *J. Phys. Chem. C* **122**, 11414–11421 (2018).
16. R. K. Pandey, H. Kakehashi, H. Nakanishi, S. Soh, Correlating material transfer and charge transfer in contact electrification. *J. Phys. Chem. C* **122**, 16154–16160 (2018).
17. J. Haeberle, A. Schella, M. Sperl, M. Schröter, P. Born, Double origin of stochastic granular tribocharging. *Soft Matter* **14**, 4987–4995 (2018).
18. A. E. Wang, P. S. Gil, M. Holonga, Z. Yavuz, H. T. Baytekin, R. M. Sankaran, D. J. Lacks, Dependence of triboelectric charging behavior on material microstructure. *Phys. Rev. Mater.* **1**, 035605 (2017).
19. D. J. Lacks, The unpredictability of electrostatic charging. *Angew. Chem. Int. Ed.* **51**, 6822–6823 (2012).
20. D. J. Lacks, T. Shinbrot, Long-standing and unresolved issues in triboelectric charging. *Nat. Rev. Chem.* **3**, 465–476 (2019).
21. F.-R. Fan, Z.-Q. Tian, Z. L. Wang, Flexible triboelectric generator. *Nano Energy* **1**, 328–334 (2012).
22. Z. L. Wang, L. Lin, J. Chen, S. Niu, Y. Zi, *Triboelectric Nanogenerators* (Springer International Publishing AG Switzerland, 2016).
23. C. Xu, Y. Zi, A. C. Wang, H. Zou, Y. Dai, X. He, P. Wang, Y.-C. Wang, P. Feng, D. Li, Z. L. Wang, On the electron-transfer mechanism in the contact-electrification effect. *Adv. Mater.* **30**, 1706790 (2018).
24. C. Xu, A. C. Wang, H. Zou, B. Zhang, C. Zhang, Y. Zi, L. Pan, P. Wang, P. Feng, Z. Lin, Z. L. Wang, Raising the working temperature of a triboelectric nanogenerator by quenching down electron thermionic emission in contact-electrification. *Adv. Mater.* **30**, 1803968 (2018).
25. C. Xu, B. Zhang, A. C. Wang, H. Zou, G. Liu, W. Ding, C. Wu, M. Ma, P. Peng, Z. Lin, Z. L. Wang, Contact-electrification between two identical materials: Curvature effect. *ACS Nano* **13**, 2034–2041 (2019).
26. C. Xu, B. Zhang, A. C. Wang, W. Cai, Y. Zi, P. Feng, Z. L. Wang, Effects of metal work function and contact potential difference on electron thermionic emission in contact electrification. *Adv. Funct. Mater.* **29**, 1903142 (2019).
27. S. Lin, C. Xu, L. Xu, Z. L. Wang, The overlapped electron-cloud model for electron transfer in contact electrification. *Adv. Funct. Mater.* **30**, 1909724 (2020).
28. S. Lin, L. Xu, L. Zhu, X. Chen, Z. L. Wang, Electron transfer in nanoscale contact electrification: Photon excitation effect. *Adv. Mater.* **31**, 1901418 (2019).
29. S. Lin, L. Xu, W. Tang, X. Chen, Z. L. Wang, Electron transfer in nano-scale contact electrification: Atmosphere effect on the surface states of dielectrics. *Nano Energy* **65**, 103956 (2019).
30. S. Lin, M. Zheng, J. Luo, Z. L. Wang, Effects of surface functional groups on electron transfer at liquid-solid interfacial contact electrification. *ACS Nano* **14**, 10733–10741 (2020).
31. M. Zheng, S. Lin, L. Xu, L. Zhu, Z. L. Wang, Scanning probing of the tribovoltaic effect at the sliding interface of two semiconductors. *Adv. Mater.* **32**, 2000928 (2020).
32. S. Lin, L. Xu, A. Chi Wang, Z. L. Wang, Quantifying electron-transfer in liquid-solid contact electrification and the formation of electric double-layer. *Nat. Commun.* **11**, 399 (2020).
33. M. Willatzen, L. C. L. Y. Voon, Z. L. Wang, Quantum theory of contact electrification for fluids and solids. *Adv. Funct. Mater.* **30**, 1910461 (2020).
34. R. Alicki, A. Jenkins, Quantum theory of triboelectricity. *Phys. Rev. Lett.* **125**, (2020).
35. Z. L. Wang, A. C. Wang, On the origin of contact-electrification. *Mater. Today* **30**, 34–51 (2019).
36. D. O. Olawale, O. O. I. Okoli, R. S. Fontenot, W. A. Hollerman, *Triboluminescence Theory, Synthesis, and Application* (Springer International Publishing AG Switzerland, 2016).
37. C. G. Camara, J. V. Escobar, J. R. Hird, S. J. Putterman, Correlation between nanosecond x-ray flashes and stick–slip friction in peeling tape. *Nature* **455**, 1087–1089 (2008).
38. A. Kramida, Y. Ralchenko, J. Reader, NIST ASD Team, NIST Atomic Spectra Database (ver. 5.8) (National Institute of Standards and Technology, 2020).
39. E. H. Wichmann, *Quantum Physics (In SI Units), Berkeley Physics Course-Volume 4* (McGraw-Hill Asia Holdings (Singapore) PTE. LTD and China Machine, 2014).
40. G. W. F. Drake, *Atomic, Molecular, and Optical Physics Handbook* (AIP, 1996).
41. H. N. Russell, A. G. Shenstone, L. A. Turner, Report on notation for atomic spectra. *Phy. Rev.* **33**, 900–906 (1929).
42. Z. L. Wang, Triboelectric nanogenerator (TENG)—Sparking an energy and sensor revolution. *Adv Energy Mater* **10**, 2000137 (2020).
43. K. Momma, F. Izumi, VESTA 3 for three-dimensional visualization of crystal, volumetric and morphology data. *J. Appl. Cryst.* **44**, 1272–1276 (2011).
**Acknowledgments**
**Funding:** This work was supported by National Natural Science Foundation of China (grant 61804010), National Natural Science Foundation of China (grant 11704032), National Natural Science Foundation of China (grant 51432005), National Key R&D Project from Minister of Science and Technology (2016YFA0202704), and Beijing Municipal Science and Technology Commission (Z171100000317001, Z171100002017017, and Y3993113DF). **Author contributions:** Conceptualization: D.L. and C.X. conceived the idea under the supervision of Z.L.W. Methodology: D.L. and C.X. designed the experiments. Investigation: D.L., Y.L., W.C., and Y.Z. performed the measurements and did data analysis. Supervision: Z.L.W. Writing (original draft): D.L. and Z.L.W. discussed the data and prepared the manuscript, with input from all authors. **Competing interests:** The authors declare that they have no competing interests. **Data and materials availability:** All data needed to evaluate the conclusions in the paper are present in the paper and/or the Supplementary Materials.

Submitted 16 April 2021
Accepted 4 August 2021
Published 24 September 2021
10.1126/sciadv.abj0349

**Citation:** D. Li, C. Xu, Y. Liao, W. Cai, Y. Zhu, Z. L. Wang, Interface inter-atomic electron-transition induced photon emission in contact-electrification. *Sci. Adv.* **7**, eabj0349 (2021).
Downloaded from https://www.science.org on September 24, 2021Li *et al.*, *Sci. Adv.* 2021; **7** : eabj0349    24 September 2021    8 of 8

# Interface inter-atomic electron-transition induced photon emission in contact-electrification

Ding LiCheng XuYanjun LiaoWenzhe CaiYongqiao ZhuZhong Lin Wang

**View the article online**
https://www.science.org/doi/10.1126/sciadv.abj0349
**Permissions**
https://www.science.org/help/reprints-and-permissions